\documentclass[11pt]{article}
\newcommand{\Comment}[1]{{}}
\usepackage[textwidth = 430 pt, textheight = 630 pt]{geometry}
\usepackage{amssymb,euscript,amsmath,amsfonts,pdfsync}

\usepackage{color}
\definecolor{MyDarkBlue}{rgb}{0.15,0.15,0.45}
\usepackage[linktocpage=true]{hyperref}
\hypersetup{
colorlinks=true,
citecolor=MyDarkBlue,
linkcolor=MyDarkBlue,
urlcolor=MyDarkBlue,
pdfauthor={Neil Lambert, Paul Richmond},
pdftitle={(2,0) Supersymmetry and the Light-Cone Description of M5-branes},
pdfsubject={hep-th}
}

\usepackage[numbers,sort&compress]{natbib}
\usepackage{hypernat}

\newcommand\ignore[1]{}
\def\one{{\,\hbox{1\kern-.8mm l}}}

\newcommand{\Cset}{{\,\,{{{^{_{\pmb{\mid}}}}\kern-.45em{\mathrm C}}}}}

\newcommand{\be}{\begin{equation}}
\newcommand{\ee}{\end{equation}}
\newcommand{\bea}{\begin{eqnarray}}
\newcommand{\eea}{\end{eqnarray}}

\begin{document}

\renewcommand{\thefootnote}{\fnsymbol{footnote}}

\rightline{CERN-PH-TH/2011-214}
\rightline{KCL-MTH-11-16}

   \vspace{1.8truecm}

\vspace{15pt}

\centerline{\LARGE \bf {\sc $(2,0)$ Supersymmetry}} \vspace{.5cm} \centerline{\LARGE \bf {\sc  And The Light-Cone Description of M5-branes}} \vspace{2truecm} \thispagestyle{empty} \centerline{
    {\large {\bf {\sc N.~Lambert${}^{\,a,}$}}}\footnote{On leave of absence from King's College London.}$^,$\footnote{E-mail address: \href{mailto:neil.lambert@cern.ch}{\tt neil.lambert@cern.ch}} {and}
    {\large {\bf{\sc P. Richmond${}^{\,b,}$}}}\footnote{E-mail address:
                                 \href{mailto:paul.richmond@kcl.ac.uk}{\tt paul.richmond@kcl.ac.uk}}         }

\vspace{1cm}
\centerline{${}^a${\it Theory Division, CERN}}
\centerline{{\it 1211 Geneva 23, Switzerland}}
\vspace{.8cm}
\centerline{${}^b${\it Department of Mathematics, King's College London}}
\centerline{{\it The Strand, London WC2R 2LS, UK}}

\bigskip
\begin{center}
{\bf {\sc Abstract}}
\end{center}
In    1007.2982 a novel system of equations which propagate in one null and four space directions were obtained as the on-shell conditions for the six-dimensional  $(2,0)$ superalgebra.  In this paper we show how this  system reduces to one-dimensional motion on instanton moduli space. Quantization leads to the previous light-cone proposal of the $(2,0)$ theory, generalized to include a  potential that arises on the Coulomb branch as well as couplings to background gauge and self-dual two-form fields.

\newpage

\section{\sl Introduction}

M-theory is well-known but not well-understood. It arises as an umbrella theory that unifies the various perturbative string theories into a single non-perturbative theory. In its strong coupling phase M-theory does not possess string states but rather M2-branes and M5-branes. The M2-branes are now relatively well understood and described by three-dimensional conformal Chern-Simons-Matter theories with  16 (BLG model \cite{Bagger:2006sk}\cite{Gustavsson:2007vu}) or 12 (ABJM models \cite{Aharony:2008ug}) manifest supersymmetries. The M5-brane however remains very mysterious. Its worldvolume description arises from a six-dimensional conformal theory with $(2,0)$ supersymmetry. Unlike the case of the three-dimensional worldvolume theories of M2-branes very little is known about six-dimensional UV complete quantum field theories, let alone those with maximal supersymmetry.

There have been several attempts to understand the $(2,0)$ theory in the literature. Some time ago a light-cone formulation was proposed in \cite{Aharony:1997th}, for the case of light-like compactification of the M5-brane as well as related constructions from Matrix Theory \cite{Berkooz:1996is}\cite{Maldacena:2002rb}. In addition a four-dimensional  `deconstruction' was presented in \cite{ArkaniHamed:2001ie}. More recently it has been suggested that maximally supersymmetric five-dimensional Yang-Mills theory can be used to define the $(2,0)$ theory in the case of a space-like compactification,  including all the Kaluza-Klein modes \cite{Douglas:2010iu}\cite{Lambert:2010iw}.
Other, even more recent, discussions on formulating the dynamics of the $(2,0)$ theory are \cite{Ho:2011ni}\cite{Samtleben:2011fj}\cite{Chu:2011fd}\cite{Bolognesi:2011nh}.

An attempt to shed some light on the mysterious $(2,0)$ system and the M5-brane was presented in \cite{Lambert:2010wm}. There a non-Abelian system of equations of motion were derived which provide a representation of the $(2,0)$ superalgebra. The construction involved a new field $C^\mu_a$ however the on-shell constraints force $C^\mu_a$ to be constant and furthermore set all derivatives of the non-Abelian fields to zero along the direction of $C^\mu_a$. Thus,  although the system is formally six-dimensionally Lorentz invariant, its non-trivial dynamics are five-dimensional. Choosing a space-like vacuum expectation value for $C^\mu_a$ leads to five-dimensional maximally supersymmetric Yang-Mills. Indeed  a closely related system can essentially be reverse-engineered directly from maximally supersymmetric five-dimensional Yang-Mills \cite{Singh:2011id}. Nevertheless there have been some encouraging signs that this formalism is capable of describing various branes in string theory and M-theory \cite{Kawamoto:2011ab}\cite{Honma:2011br}. In this paper we wish to study the system of equations obtained in \cite{Lambert:2010wm} in the case where the auxiliary field $C^\mu_a$ is null.

The rest of this paper is organized as follows. In section two we review the $(2,0)$ system constructed in \cite{Lambert:2010wm}. We also determine the conserved energy momentum tensor, supercharge and compute the superalgebra including the central charges. In section three we consider in detail the resulting dynamical system when the auxiliary vector field $C^\mu_a$ has a null vacuum expectation value. This leads to a curious system of equations with 16 supersymmetries and an $SO(5)$ R-symmetry that propagate in one null and four space directions.  We show how the equations reduce to motion on instanton moduli space, where the instanton number is the null momentum parallel to $C^\mu_a$. We then quantize the system by using the other null momentum generator as a Hamiltonian. This leads directly to the light-cone quantization proposal of the $(2,0)$ theory proposed in
\cite{Aharony:1997th,Aharony:1997th,Aharony:1997an}, generalized to include a potential when the scalars   have a vacuum expectation value and also  couplings to background gauge  and self-dual two-form fields. In section four we end with our conclusions.

\section{\sl A Non-Abelian $(2,0)$ Supersymmetry}

Let us start by reviewing the construction of \cite{Lambert:2010wm}. The fields consist of 5 scalars $X^I_a$, a sixteen-component Fermion $\psi_a$ which satisfies $\Gamma_{012345}\psi_a=-\psi_a$,  a gauge field $\tilde A_\mu{}^a{}_b$, a vector $C^\mu_a$ and a self-dual three-form $H_{\mu\nu\lambda\ a}$:
\begin{align}
H_{\mu\nu\lambda \; a} &= \frac{1}{3!}\epsilon_{\mu\nu\lambda\tau\sigma\rho}H^{\tau\sigma\rho}{}_a\; .
\end{align}
 Here the index $a$ refers to the fact that the fields take values in a 3-algebra with structure constants $f^{cdb}{}_a$ which are totally anti-symmetric (when all indices are raised) and satisfy the fundamental identity
 \begin{align}
f^{efg}{}_df^{abc}{}_g = f^{efa}{}_gf^{gbc}{}_d+f^{efb}{}_gf^{agc}{}_d+f^{efc}{}_gf^{abg}{}_d\ .
\end{align}

The supersymmetry transformations are:
\begin{align}
\delta X^I_a &= i \bar{\epsilon} \Gamma^I \psi_a \\
\delta \psi_a &= \Gamma^\mu \Gamma^I \epsilon D_\mu X^I_a + \frac{1}{3!} \frac{1}{2} \Gamma_{\mu \nu \lambda} \epsilon H^{\mu \nu \lambda}_a - \frac{1}{2} \Gamma_\lambda \Gamma^{IJ} \epsilon C^\lambda_b X^I_c X^J_d f^{cdb}{}_a \\
\delta H_{\mu \nu \lambda \; a} &= 3i \bar{\epsilon } \Gamma_{[\mu \nu} D_{\lambda]} \psi_a + i \bar{\epsilon} \Gamma^I \Gamma_{\mu \nu \lambda \kappa} C^\kappa_b X^I_c \psi_d f^{cdb}{}_a \\
\delta \tilde{A}_\mu{}^b{}_a &= i \bar{\epsilon} \Gamma_{\mu \lambda} C^\lambda_c \psi_d f^{cdb}{}_a \\
\delta C^\mu_a &= 0 \, .
\end{align}
These transformations close on-shell. In particular the equations of motion are \cite{Lambert:2010wm}\footnote{Note the corrected sign in the scalar equation of motion, first pointed out in \cite{Kawamoto:2011ab}, as well the corrected sign in the gauge field equation of motion.}
\begin{align}
0 &= \Gamma^\mu D_\mu\psi_a+X^I_c C^\nu_b \Gamma_\nu\Gamma^I\psi_d f^{cdb}{}_a \\
0 &=D^2 X_a^I -\frac{i}{2}\bar\psi_c C^\nu_b \Gamma_\nu\Gamma^I \psi_d f^{cdb}{}_a + C^\nu_b C_{\nu g} X^J_c X^J_e X^I_f f^{efg}{}_{d}f^{cdb}{}_a \\
0 &= D_{[\mu}H_{\nu\lambda\rho]\;a}+\frac{1}{4}\epsilon_{\mu\nu\lambda\rho\sigma\tau}C^\sigma_b X^I_c D^\tau X^I_d f^{cdb}{}_a + \frac{i}{8}\epsilon_{\mu\nu\lambda\rho\sigma\tau}C^\sigma_b \bar\psi_c \Gamma^\tau \psi_d f^{cdb}{}_a \\
0&= \tilde F_{\mu\nu}{}^b{}_a + C^\lambda_c H_{\mu\nu\lambda\; d}f^{cdb}{}_a \\
0 &= D_\mu C^\nu_a = C^\mu_c C^\nu_d f^{bcd}{}_a \\
\label{pconstraint}0 &= C^\rho_c D_\rho X^I_d f^{cdb}{}_a = C^\rho_c D_\rho \psi_d f^{cdb}{}_a =C^\rho_c D_\rho H_{\mu\nu\lambda\; a} f^{cdb}{}_a\ .
\end{align}
In all these equations $\tilde F_{\mu\nu}{}^b{}_a$ is the field strength of the gauge connection $\tilde A_\mu{}^b{}_a$ which appears in the covariant derivative $D_\mu$ which acts as, for example,  $D_\mu X^I_a = \partial_\mu X^I_a - \tilde A_\mu{}^b{}_aX^I_b$.

Note that the second to last equation implies that $C^\mu_a$ is constant and hence selects a preferred direction spacetime and in the 3-algebra. The final equations imply that the non-Abelian components of the fields can only propagate in the five dimensions orthogonal to $C^\mu_a$.

Next it will be useful to construct the conserved currents of this theory. In particular we look for an energy-momentum tensor $T_{\mu\nu}$ as well as a supercurrent $J^\mu$.
Simple trial and error leads to the following expressions:
\begin{align}
\nonumber
T_{\mu \nu} =& D_\mu X^I_a D_\nu X^{Ia} - \frac{1}{2} \eta_{\mu \nu} D_\lambda X^I_a D^\lambda X^{Ia} \\
\nonumber
	&+ \frac{1}{4} \eta_{\mu \nu} C^\lambda_b X^I_a X^J_c C_{\lambda g} X^I_f X^J_e f^{cdba} f^{efg}{}_d + \frac{1}{4} H_{\mu \lambda \rho \; a} H_{\nu}{}^{\lambda \rho \; a} \\
	&- \frac{i}{2} \bar{\psi}_a \Gamma_\mu D_\nu \psi^a + \frac{i}{2} \eta_{\mu \nu} \bar{\psi}_a \Gamma^\lambda D_\lambda \psi^a - \frac{i}{2} \eta_{\mu \nu} \bar{\psi}_a C^\lambda_b X^I_c \Gamma_\lambda \Gamma^I \psi_d f^{abcd} \, \\
	\nonumber \\
J^\mu =& \frac{1}{2} \frac{1}{3!} H_{\nu \lambda \rho \; a} \Gamma^{\nu \lambda \rho} \Gamma^\mu \psi^a - D_\nu X^I_a \Gamma^\nu \Gamma^I \Gamma^\mu \psi^a - \frac{1}{2} C^\nu_b X^I_c X^J_d \Gamma_\nu \Gamma^{IJ} \Gamma^\mu \psi^a f^{bcd}{}_a \, .
\end{align}
In the Abelian case this agrees with the linearised form of the energy-momentum tensor derived in \cite{Barwald:1999jh}. The associated conserved charges are
\begin{align}
P_\mu = \int d^5 x \ T_{\mu 0}\ ,\qquad \qquad Q = \int d^5 x \ J^0\ ,
\end{align}
where the integrals are over the spatial coordinates, corresponding to the momentum and supercharge respectively.

The superalgebra of the $(2,0)$-theory can then be deduced by evaluating $\delta J^0 = \delta _\epsilon J^{0\alpha} \epsilon_\alpha$ {\it viz}:
\begin{align}
\nonumber
\{ Q_\alpha , Q_\beta \} =& - \int d^5x \ ( \delta_\epsilon J^0 C^{-1} )_{\alpha \beta} \\
=&-2 ( \Gamma^\mu C^{-1} )_{\alpha \beta} P_\mu + ( \Gamma^\mu \Gamma^I C^{-1} )_{\alpha \beta} Z_\mu^{I} + ( \Gamma^{\mu \nu \lambda} \Gamma^{IJ} C^{-1} )_{\alpha \beta} Z_{\mu \nu \lambda}^{IJ} \, .
\end{align}
The central charges we obtain in this way are (in the case of vanishing Fermions):
\begin{align}
Z_0^{I} &= \int  d^5x \ 4 C^0_{b} X^I_c X^J_d D^0 X^J_a f^{cdba}  \\
\nonumber Z_i^I &= \int d^5x \ \left( H^{0}{}_{j i \; a} D^j X^{I a}  +  \frac{1}{6} \partial_{j} ( H_{klm \; a} X^{Ia} \varepsilon_{0ijklm} )\right. \\
& \left. \hskip2cm + C_{i b} X^I_c X^J_d D^0 X^J_a f^{cdba}- 2 C^0_{b} X^I_c X^J_d D_i X^I_a f^{cdba} \right) \\
Z_{0 i j}^{IJ} &= \int d^5x \  \left( \frac{1}{2} H_{0 i j \; a} C^0_b X^I_c X^J_d f^{cdba} - \partial_i ( X^I_a D_j X^{Ja} ) \right) \\
Z_{k l m}^{IJ} &= \int d^5x \  \left( \frac{1}{12} H_{k l m \; a} C^0_b X^I_c X^J_d f^{cdba} + \frac{1}{36} \partial_i ( C_{j b} X^K_c X^L_d X^M_a f^{cdba} \varepsilon_{0ijklm} \varepsilon^{IJKLM} ) \right) \, ,
\end{align}
where here, in this section, $i,j=1,2,3,4,5$.

\section{\sl Null Reduction}

Next we wish to consider the above system of equations for the special case where $C^\mu_a$ is a null vector:
\begin{equation}
C^\mu_a = \frac{g^2}{\sqrt{2}}(\delta^\mu_0+\delta^\mu_5)\delta_a^*
\end{equation}
where $g^2$ has dimensions of length and $*$ denotes some preferred direction in the 3-algebra. We choose to go to light-cone coordinates {\it i.e.}
$x^\mu = ( x^{+} , x^{-} , x^i )$ where
\begin{equation}
x^{-} = \frac{1}{\sqrt{2}} ( x^0 - x^5 )\ ,\qquad x^{+}  = \frac{1}{\sqrt{2}} (x^0 + x^5 )\ ,
\end{equation}
so that $C^\mu_a = g^2\delta^\mu_+\delta_a^*$. Note that for the rest of this paper we have $i,j=1,2,3,4$ (rather than $i,j=1,2,3,4,5$ that was used in the previous section). The constraint (\ref{pconstraint}) now tells us that $D_{+}$ vanishes on all the fields. Furthermore $\tilde F_{i+}{}^b{}_a=0$ so $\tilde A_+{}^b{}_a$ is a flat connection and can be set to zero (at least locally). Thus the fields are essentially just functions of $x^i$ and $x^-$. Here we wish to view these equations of motion as a dynamical system where $x^-$ plays the role of time.

Let us now give the equations of motion that follow from the choice $C^\mu_a = g^2 \delta^\mu_+\delta_a^*$. Fixing the element $T^*$ in the 3-algebra means that the remaining generators behave as an ordinary Lie-algebra with Lie bracket:
\begin{equation}
i [T^c,T^d]= [T^*,T^c,T^d]=f^{*cd}{}_a T^a\ .
\end{equation}
The components of the fields along the $*$ direction in the 3-algebra decouple and behave as a free six-dimensional tensor multiplet and for the rest of this paper we simply discard them. Alternatively one could have started from a non-Abelian $(2,0)$ system where the C-field does not take values in the algebra, {\it i.e.} $C^\mu$ instead of $C^\mu_a$, as in the construction of \cite{Singh:2011id}.

For the sake of clarity we will use a notation whereby all the fields are taken to be Lie-algebra valued: {\it e.g.} $X^I = \sum_{a\ne *}X_a^I T^a$, and the $a$ index is dropped. We also note that the gauge field $\tilde A_\mu{}^b{}_a$ and field strength $\tilde F_{\mu\nu}{}^b{}_a$ also take values in the Lie-algebra and act on the other fields through the commutator. Therefore we drop the $a,b$ indices and tilde on these fields in what follows.

In the $(x^+,x^-,x^i)$ coordinates self-duality of $H_{\mu\nu\lambda}$ implies that $F_{ij} = -g^2 H_{ij+}$ is anti-self-dual, $G_{ij} = -g^2 H_{ij-}$ is self-dual and
\begin{equation}
H_{ijk} = g^{-2} \epsilon_{ijkl}F^l{}_-\ .
\end{equation}
Noting that the constraint implies that only the derivatives $D_-$ and $D_i$ are non-vanishing we find the remaining equations of motion can be written as
\begin{align}
0 &= \Gamma^- D_-\psi +\Gamma^iD_i\psi+ig^2[X^I  ,\Gamma_+\Gamma^I\psi]\\
0 &=D_iD^i X^I +\frac{g^2}{2}[\bar\psi, \Gamma_+\Gamma^I \psi] \\
0 &= D^iF_{i-} + \frac{g^4}{2} [\bar\psi, \Gamma_+ \psi]  \\
\label{secondorder}0 &= D_-F_{i-}-D^jG_{ij} - ig^4[X^I, D_i X^I] + \frac{g^4}{2} [\bar\psi, \Gamma_i \psi]  \\
0 &= D_{[i}F_{j-]}\ .
\end{align}
One sees that the final equation is just the Bianchi identity and automatically satisfied.

Our strategy now is to solve as many of the equations of motion as possible. We will do this by setting the Fermions to zero with the understanding that the supersymmetry can be used to generate Fermionic solutions. We will see that all but the second order equation (\ref{secondorder}) can be solved and reduced to ADHM data.

To continue we first observe that the gauge field $A_i$ is determined by the ADHM construction \cite{Atiyah:1978ri}. Thus the degrees of freedom of the gauge field are reduced to the finite dimensional instanton moduli space with local coordinates $m^\alpha$. Note that although $G_{ij}$
is self-dual it has no interpretation as the field strength of $A_i$. Therefore $G_{ij}$ is not necessarily the field strength of a gauge field and one cannot solve for it using the ADHM construction. In fact $G_{ij}$ behaves as a non-dynamical background field since its $D_-$ derivative never appears.

With vanishing Fermions the scalar equation of motion is just $D_iD^i X^I=0$. It is easy to see that there is a unique solution to this equation for any given asymptotic value of $X^I$. In addition for an instanton background there exists smooth solutions. Thus $X^I$ is uniquely determined in terms of the ADHM data of the gauge field $ A_i$ and its asymptotic value:
\begin{equation}
X^I = v^I+{\cal O}\left(\frac{1}{|x|^2}\right)\ ,
\end{equation}
where $v^I$ is an element of the Lie-algebra.

Next we consider the equation $D^i F_{i-}=0$. In terms of gauge fields this is
\begin{equation}
D^iD_i A_- - D^i\partial_-A_i=0\ .
\end{equation}
To solve this equation we need to recall some facts about instanton moduli space, for reviews see \cite{Vandoren:2008xg},\cite{Tong:2005un}. In particular the instanton equations are
\begin{equation}
F_{ij} =-\frac{1}{2}\varepsilon_{ijkl}F^{kl}\ .
\end{equation}
Moduli correspond to infinitesimal changes to the gauge fields that preserve this condition:
\begin{equation}
D_i \delta A_{j} - D_j\delta A_{i} = -\varepsilon_{ijkl}D^k\delta A^l\ .
\end{equation}
However gauge transformations $\delta A_{i}= D_i\omega$ will clearly solve these equations and we do not wish to include them in the moduli. To exclude them we require that $\delta A_{i}$ is orthogonal to all gauge modes:
\begin{equation}
{\rm Tr}\int d^4 x \ \delta A_{i} D^i\omega =0\ .
\end{equation}
Integrating by parts, and requiring that $\omega=0$ at infinity,  shows that we therefore impose the gauge fixing condition
\begin{equation}
D^i\delta A_{i}=0\ .
\end{equation}
We have seen that the solution to the equations of motion requires that $A_{i}$ has anti-self-dual field strength. Therefore the $x^-$ dependence comes entirely through the dependence of the moduli on $x^-$ and hence we conclude that
\begin{equation}\label{gauge}
D^i\partial_-  A_{i}=0\ ,
\end{equation}
with $\partial_-A_i = \frac{\partial A_i}{\partial m^\alpha} \partial_-m^\alpha+D_i\omega$ where $\omega$ is chosen to ensure that (\ref{gauge}) is satisfied.
Thus the $D^i F_{i-} =0 $ equation simply becomes $D^iD_i A_-=0$. This is the same as the  $X^I$ equation and so $A_-$ is also determined in terms of ADHM data and its asymptotic value:
\begin{equation}
A_- = w+{\cal O}\left(\frac{1}{|x|^2}\right)\ ,
\end{equation}
where $w$ is an element of the Lie-algebra.

We are now left with just one equation which is second order in $x^-$:
\begin{equation}\label{dd}
D_-F_{i-}-D^jG_{ij} - ig^4[X^I, D_i X^I]=0\ .
\end{equation}
However as we mentioned above we do not aim to solve this equation - which would amount to a complete solution to all the classical field equations. Rather we now wish to quantize the classical field configurations that we have constructed and use the momentum generator along $x^-$ as the Hamiltonian.

\subsection{\sl Conserved Charges}

To proceed we note that we need to use a slightly different definition of the conserved charge. In particular the problem with the standard definition given in section 2 is that the integral over all space includes an integral over $x^5$. However one can simply change integration variable from $x^5$ to $x^-$ so that the integral is over all the coordinates. The resulting conserved charge is therefore constant not for dynamical reasons but because we have integrated over all the coordinates upon which the fields depend.

On the other hand we can consider
\begin{align}
{\cal P}_\mu = g^2\int d^4 x \ T_{\mu +}\ ,\qquad \qquad {\cal Q} = g^2\int d^4 x \ J^-\ ,
\end{align}
where we have included a factor of $g^2$ to ensure that they have the canonical dimensions.
Since $D_+=0$, ${\cal P}_\mu$ and ${\cal Q}$ are conserved in the sense that $\partial_-{\cal P}_\mu=\partial_-{\cal Q}=0$. Note that this assumes that the fields vanish sufficiently quickly at infinity so that the boundary terms in the integrals can be discarded. In particular conservation of ${\cal Q}$ requires that $D_-X^I$ and $[X^I,X^J]\to 0$ as $x^i\to\infty$. Therefore, in this paper, in order to obtain conserved charges that can be used to define the quantum theory we assume that
\begin{equation}\label{vbdry}
[v^I,v^J]=[v^I,w]=0\ .
\end{equation}
{\it i.e.} we require that the scalar fields and gauge field are in a vacuum configuration at infinity.
More explicitly these expressions are (in the case of vanishing Fermions):
\begin{align}
{\cal P}_- =& {\rm Tr}\int d^4 x \   \frac{1}{2g^2} F_{i-} F^i{}_- +\frac{g^2}{2} D_i X^ID^i X^{I} \\
{\cal P}_+ =& -\frac{1}{8g^2}{\rm Tr}\int d^4 x \    \varepsilon^{ijkl}F_{ij} F_{kl}\\
{\cal P}_i =& \frac{1}{2g^2}{\rm Tr}\int d^4 x \   F_{ij} F_-{}^j \\
{\cal Q} =& {\rm Tr}\int d^4 x\   F_{i-} \Gamma^i \Gamma^- \psi - \frac{1}{4} F_{ij}\Gamma^{i j} \Gamma^+ \Gamma^- \psi + g^2 D_i X^J \Gamma^J \Gamma^i \Gamma^- \psi \, .
\end{align}
Note that ${\cal P}_+ = -4\pi^2 g^{-2}k $, where $k$ is the instanton number. Thus the ${\cal P}_+$ eigenvalues are discrete. Physically we interpret this a arising because the $x^+$ direction is resticted to lie on a circle with radius $R=g^2/4\pi^2$.

We can further decompose $\cal Q={\cal Q}_++{\cal Q}_-$ where $\Gamma_{-+}{\cal Q}_\pm=\pm{\cal Q}_\pm$. In this case the superalgebra becomes
\begin{align}
\{ \cal Q_{-\alpha} , \cal Q_{-\beta }\}
=& - 2 {\cal P}_- ( \Gamma^- C^{-1} )_{\alpha \beta} + {\cal Z}^I_+ ( \Gamma^- \Gamma^I C^{-1} )_{\alpha \beta} + {\cal Z}^{IJ}_{ij+} ( \Gamma^{ij} \Gamma^- \Gamma^{IJ} C^{-1})_{\alpha \beta}\\
\nonumber \\
\{ \cal Q_{+\alpha} , \cal Q_{+\beta } \}
=& - 2 {\cal P}_+ ( \Gamma^+ C^{-1} )_{\alpha \beta} \\
\nonumber \\
\{ \cal Q_{-\alpha} , \cal Q_{+\beta } \}
=& - 2 {\cal P}_i ( \Gamma^i C^{-1} )_{\alpha \beta} + {\cal Z}^I_i  (\Gamma^i \Gamma^I C^{-1} )_{\alpha \beta} \, ,
\end{align}
where $C=\Gamma_0$ is the charge conjugation matrix and the central charges are
\begin{align}
{\cal Z}^I_+=& -2{\rm Tr}\int d^4 x \  F_{-i} D^i X^I \\
{\cal Z}_i^I=& -{\rm Tr}\int d^4 x \  G_{ij} D^j X^I\\
{\cal Z}^{IJ}_{ij+}=& -g^2{\rm Tr}\int d^4 x \  D_{[i} X^I D_{j]} X^J\ .
\end{align}
Note that although there are 16 supersymmetry charges only 8 of them ($\cal Q_-$) have a non-trivial relation with $\cal P_-$. This is a well-known feature of light-cone gauge ({\it c.f.} the Green-Schwarz superstring).  Furthermore any state with a non-vanishing ${\cal P}_+$ must break the ${\cal Q}_+$ supersymmetries. 

We also see that $G_{ij}$ only appears through its contribution to the central charge ${\cal Z}_i^I$. Here we take it to be a background, non-dynamical field, in which case one only seems to obtain a conserved quantity in the case that $D^jG_{ij}=0$, so that it decouples from (\ref{dd}). In this case  ${\cal Z}_i^I$ is simply a boundary term depending on $v^I$ and $G_{ij}$.

Thus, to summarize, we impose the constraints $D^iG_{ij}=[v^I,v^J]=[v^I,w]=0$ on the fields to ensure that there  charges given above are well-defined and conserved. This is necessary in our treatment since we will ultimately quantize the theory and use the Hamiltonian as the generator of time evolution through a Schr\"odinger equation.

\section{\sl Quantization}

We have seen above that the classical equations of motion can be solved up to a single second order evolution. We have also constructed the conserved momentum and central charges in the $(2,0)$ algebra. In this section, rather than solve the second order classical evolution equation we instead wish to quantize the system using ${\cal P}_-$ as the Hamiltonian. In particular we see that it can be written as
\begin{equation}
{\cal P}_- = \frac{1}{2g^2}{\rm Tr}\int d^4 x \ \partial_-A_i\partial_-A^i - 2\partial_-A_i D^i A_- + D_iA_-D^iA_-+g^4D_iX^ID^iX^I   \ .
\end{equation}
The first term gives the kinetic energy and can be expressed in terms of the metric $g_{\alpha\beta}$ on instanton moduli space defined by
\begin{equation}
{\rm Tr}\int d^4 x \ \delta A_i \delta A^i = g_{\alpha\beta}\delta m^{\alpha}\delta m^\beta \, .
\end{equation}
Here $\delta A_i = \partial A_i/\partial m^\alpha \delta m^\alpha + D_i\delta \omega$,  with $\delta\omega$ is the gauge transformation required to preserve $D^i\delta A_i=0$.

Next we have a term that is linear in time derivatives:
\begin{equation}
{\rm Tr}\int d^4 x \ \partial_-A_i D^i A_- = {\rm Tr}\oint \partial_-A_r w = L_\alpha \dot m^\alpha\ .
\end{equation}
where $r$ is the radial normal direction to the sphere at infinity,  $\dot m^\alpha = \partial_- m^\alpha$  and $L_\alpha$ is a vector field on the instanton moduli space. We note that it is proportional to $w$, {\it i.e.} it is determined by the vacuum expectation value of $A_-$, and can be viewed as a background gauge field.

 The last two terms can be written as a boundary integral and contribute to the potential. Thus we find that the Hamiltonian is
 \begin{equation}
{\cal P}_-= \frac{1}{2g^2}g_{\alpha\beta} (\dot m^\alpha - L^\alpha) (\dot m^\beta-L^\beta )+V\ ,
 \end{equation}
 where
 \begin{equation}\label{V}
 V  = - \frac{1}{2g^2} g_{\alpha\beta}L^\alpha L^\beta+\frac{1}{2g^2}{\rm Tr}\oint g^{4}X^ID_rX^I+A_-D_rA_-\  .
 \end{equation}

 For $w=0$ this Hamiltonian has appeared before \cite{Lambert:1999ua}  and is known to admit 8 supersymmetries, which correspond to the ${\cal Q}_-$ here. In particular it was shown that
 \begin{equation}
 V = \frac{g^2}{2} g_{\alpha\beta} K^\alpha K^\beta\ ,
 \end{equation}
 where $K^\alpha$ is a tri-holomorphic Killing vector on the instanton moduli space which
 can be expressed purely in terms of the asymptotic values of $X^I$ and the ADHM data  \cite{Lambert:1999ua} . By construction the Hamiltonian is also invariant under 8 supersymmetries when $w\ne 0$.

The next step is to decide on a momentum conjugate to the moduli coordinates $m^\alpha$. The obvious choice is
\begin{equation}
p_\alpha = g_{\alpha\beta}  \dot m^\beta \ .
\end{equation}
An alternative quantization could be $p_\alpha = g_{\alpha\beta} (\dot m^\beta-L^\beta)$ however since $L^\alpha$ depends on $w_a$ this quantization would then differ in various sectors of the theory.
It would be interesting to obtain a symplectic structure on the entire $(2,0)$ system that leads to this.
Quantization is now straightforward and we just consider wavefunctions $\Psi(m^\alpha,x^-)$ and define
\begin{equation}
\hat p_\alpha\Psi = -i\frac{\partial \Psi}{\partial m^\alpha}\, , \qquad \qquad \hat m^\alpha \Psi = m^\alpha \Psi \, ,
\end{equation}
where a hat denotes the quantum operator. 

There is  one issue that requires some discussion, namely the moduli space generically contains singularities where the instantons shrink to zero size. These are not curvature singularities but rather more like orbifold singularities. Thus we should either seek to remove them or simply come up with  a suitable prescription on the behaviour of the wavefunction at the singularities. Methods for pursuing the first approach were considered in \cite{Aharony:1997th}. For the second approach one could simply assume that physical wavefunctions need to be even under the orbifold action at each singularity.

\subsection{\sl One Instanton Example}

For concreteness we now give the expressions above for the case of a single instanton {\it i.e.}:
\begin{equation}
{\cal P}_+=-4\pi^2/g^2
 \end{equation}
 with gauge group $SU(2)$, including all the moduli. In this case we have ($\eta^a_{ij}$ are the self-dual 't Hooft matrices)
\begin{eqnarray}
A_i &=& \frac{1}{(x-y)^2}\frac{\rho^2}{(x-y)^2+\rho^2}\eta^a_{ij}(x-y)^j U\sigma_a U^{-1}\\
X^I &=& \frac{(x-y)^2}{(x-y)^2+\rho^2}{v^I_aU\sigma_a U^{-1}}\\
 A_- &=&\frac{(x-y)^2}{(x-y)^2+\rho^2}{w_aU\sigma_a U^{-1} }\ .
 \end{eqnarray}
Here there are eight moduli represented by the instanton size $\rho$, position $y^i$ and gauge embedding $U\in SU(2)\equiv S^3$. Therefore, in total the moduli space is eight-dimensional.

Our first task is to compute the metric. To do this we note that to ensure $D^i\partial_-A_i=0$ we find that $\omega$ is given by
\begin{equation}
\omega =  \frac{1}{(x-y)^2}\frac{\rho^2}{(x-y)^2+\rho^2} \eta^a_{ij}\dot y ^i (x-y)^jU\sigma_aU^{-1}- \frac{\rho^2}{(x-y)^2+\rho^2} \dot u^a U\sigma_a U^{-1}\ ,
\end{equation}
where we have introduced
\begin{equation}
U^{-1}\dot U = i\dot u^a\sigma_a\ . 
\end{equation}
We can now compute the metric and find
\begin{equation}
ds^2 = 8\pi^2 (d\rho^2+\rho^2du^adu^a)+4\pi^2dy^kdy^k\ .
\end{equation}
This is just the flat metric on ${\mathbb R}^4\times {\mathbb R}^4$ ($u^a$ are the left-invariant $SU(2)$ forms of the unit $S^3$). However we note that, by construction, $U$ is indistinguishable from $-U$ and therefore  the actual moduli space is obtained by identifying $U\cong -U$ and hence is the quotient ${\mathbb R}^4/{\mathbb Z}_2\times {\mathbb R}^4$.

Next we evaluate
\begin{equation}\label{bt}
\oint  \partial_-A_r = \oint \frac{\partial A_r}{\partial m^\alpha} \dot m^\alpha + D_r \omega\ ,
\end{equation}
where $r$ is the normal direction to the boundary. The only contributions to this come from the ${\cal O}(1/r^3)$ term in $\partial_-A_r$.  To evaluate (\ref{bt}) one notes that  $\partial A_i /\partial y^k = {\cal O}(r^{-4})$ and, although the $\partial A_i /\partial \rho$  and $\partial A_i/\partial U$ terms are ${\cal O}(r^{-3})$, their $\partial A_r /\partial \rho$  and $\partial A_r/\partial U$ components vanish. Thus we have
\begin{equation}\label{btone}
\oint  \partial_-A_r = \oint  D_r \omega =  4\pi^2 \rho^2U\dot u^a\sigma_aU^{-1}\ ,
\end{equation}
and hence
\begin{equation}
L_\alpha \dot m^\alpha=   8\pi^2 \rho^2 w_a\dot u^a\ ,
\end{equation}
or equivalently $L^\alpha=w^a\delta_a^\alpha$. If we consider gauge transformations of the form $U(x^-)$ then $L^\alpha$ will transform as a gauge field.
For $V$ we find
\begin{equation}
V =  4\pi^2g^2 v^I_av^I_a  \rho^2\ .
\end{equation}
Note that the first and last terms in (\ref{V}) have completely cancelled each other and we expect that this is generically the case. Thus we have found that
\begin{equation}
{\cal P}_- = \frac{4\pi^2} {g^{2}}\left(\dot \rho^2 +  \rho^2(\dot u^a - w^a)(\dot u^a - w^a)+\frac{1}{2}\dot y^k\dot y^k\right)\ +  4\pi^2g^2v^I_av^I_a  \rho^2 .
\end{equation}
It is also straightforward to show that the conserved momentum is
\begin{equation}
{\cal P}_i = -2\pi^2g^{-2} \dot y_i\ .
\end{equation}
 More generally, for the case of  point-like multi-instantons   ({\it i.e.} widely separated compared to their individual scale sizes), one finds  ${\cal P}_i\sim -2\pi^2g^{-2}\sum\dot y_i$ is just the centre of mass momentum.

Let us now discuss the central charges. First consider ${\cal Z}^I_+$;
\begin{eqnarray}
\nonumber {\cal Z}^I_+ &=& - 2{\rm Tr} \int d^4x (\partial_- A_i-D_iA_-) D^iX^I\\
&=& - 2{\rm Tr} \oint (\partial_- A_r-D_rA_-)X^I\\
\nonumber &=&  -16\pi^2 \rho^2 v^I_a( \dot u^a -   w^a )\ .
\end{eqnarray}
This is  the angular momentum associated to the action of $SU(2)$ on the moduli space.

In the one-instanton case the unique solution to $D^iG_{ij}=0$ is given by \ $G_{ij} = G_0 {(x^2+\rho^2)^2}{x^{-4}}\eta^a_{ij}\sigma_a$  where $G_0$ is a constant. However conservation of ${\cal Q}$ and ${\cal P}_\mu$ requires that all fields vanish at infinity (and aren't too singular at the origin) and hence we must take $G_0=0$ so that $Z_i^I=0$. We  expect that   any states that carry $Z_i^I$ charge are string-like states extended along some direction say $x^4$. In this case
the total  ${\cal P}_+$ momentum is infinite but the ${\cal P}_+$ per unit length should be finite. Therefore the quantum mechanical system reduces to motion on the monopole moduli space determined by the Nahm construction \cite{Nahm:1979yw}.

In addition we find that  ${\cal Z}^{IJ}_{ij+}$ is given by
\begin{equation}
{\cal Z}^{IJ}_{ij+} = -2\pi^2\rho^2\epsilon^{abc}\eta^a_{ij} v^I_bv^J_c\ .
\end{equation}
However this vanishes since we demand that $[v^I,v^J]=0$ in order that $\cal Q$ is conserved. More generally we expect that any  state with non-vanishing ${\cal Z}^{IJ}_{ij+}$ should have co-dimension two, corresponding to 3-brane states of the M5-brane. In this case we need to consider states with finite  ${\cal P}_+$
per unit area and the quantum mechanical system should then be reduced to the vortex moduli space.

\section{\sl Conclusion}

In this paper we have constructed the conserved energy momentum tensor and supercurrent for the $(2,0)$ system obtained in \cite{Lambert:2010wm}.
We then considered in detail the case of a null reduction to a novel dynamical system with 16 supersymmetries and an $SO(5)$ R-symmetry in one null and four space dimensions. In particular we showed how the classical equations can be reduced to motion on the instanton moduli space. This allows us to quantize the system. In so doing we obtained the light-cone quantization proposal of  \cite{Aharony:1997th}, generalized to include a potential that arises when the scalars (or gauge field $A_-$) have a non-vanishing vacuum expectation value, corresponding to the Coulomb branch where the M5-branes are separated. We were also able to obtain expressions for the six-dimensional supersymmetry and Poincar\'e algebras in terms of ADHM data of the instanton moduli space. This clarifies the relation of the quantum mechanical system to the full six-dimensional one.

Finally it is instructive to see how the null reduction of the $(2,0)$ system above can be viewed as the limit of an infinite boost. This is in agreement with the general arguments for matrix models and light-cone quantization given in \cite{Seiberg:1997ad}. In particular let us return to the general discussion for arbitrary $C^\mu$ and set
$$
C^\mu =  \frac{g^2}{\sqrt{1+\beta^2}}(\beta\delta^\mu_0+\delta^\mu_5)\ ,
$$
where $\beta$ is real. For any $|\beta|<1$, $C^\mu$ is space-like and after a suitable Lorentz transformation could be taken to simply be $C^\mu = g^2\delta^\mu_5$ and one reproduces maximally supersymmetric five-dimensional super-Yang-Mills Theory. Taking $\beta\to \pm1$ corresponds to an infinite boost of the system along $x^5$ and leads to the null reduction we have discussed.

Let us see how this works in the $(2,0)$ system. We introduce coordinates
\begin{align}
u = \frac{x^0-\beta x^5}{\sqrt{1+\beta^2}} \, , \qquad \qquad v=\frac{x^5+\beta x^0}{\sqrt{1+\beta^2}} \, ,
\end{align}
so that $C^\mu=g^2\delta^\mu_v$ (again we are cavalier about the 3-algebra indices for the sake of clarity).
We now find that if we let
\begin{align}
F_{ij} &= -g^2 H_{ijv} \\
F_{iu} &= -g^2 H_{iuv} \\
G_{ij} &= -g^2 H_{iju} \, ,
\end{align}
then self-duality of $H$ implies that $H_{ijk} = g^{-2} \varepsilon_{ijkl} F^l{}_u$ and also:
\begin{align}
\frac{1}{2}\varepsilon_{ijkl} F^{kl} =\frac{2\beta}{1+\beta^2} F_{ij}+ \frac{1-\beta^2}{1+\beta^2} G_{ij}\ .
\end{align}
In the limit that $\beta = 1-\varepsilon$ with $\epsilon <<1$ we see that
\begin{align}
\frac{1}{2}\varepsilon_{ijkl} F^{kl} =F_{ij}+\varepsilon G_{ij}+{\cal O}(\varepsilon^2)\ ,
\end{align}
and therefore the non-self-dual part of $F_{ij}$ is boosted away. However for any $\beta \ne \pm 1$ the gauge fields are not required to be self-dual and the reduction to quantum mechanics that we found above will not occur.

In our opinion this work presents evidence that the $(2,0)$ system of  \cite{Lambert:2010wm} presents a complete Lorentz covariant picture of the M5-brane on a six-dimensional spacetime of the form ${\cal M}\times S^1$. In particular it is capable of including and interpolating between two conjectures on the dynamics of M5-branes: namely the recent suggestions that the $(2,0)$ theory on a space-like circle is precisely five-dimensional maximally supersymmetric Yang-Mills \cite{Douglas:2010iu}\cite{Lambert:2010iw} and also the older light-cone proposal of  \cite{Aharony:1997th}. In particular the latter can now be seen to arise as a space-like boost of the former in accordance with the general prescription of \cite{Seiberg:1997ad}. Nevertheless it remains to be seen if these conjectures can be made to lead to a more robust and complete description of the $(2,0)$ theory and hence the M5-brane, particularly on uncompactified spacetimes.

It could also be interesting to consider a time-like reduction. The resulting system is very similar to five-dimensional maximally supersymmetric Yang-Mills but in Euclidean signature. Although it is not clear to us what this physically means (although perhaps it could be related to the $(2,0)$ theory at finite temperature).

\section*{\sl Acknowledgements}

We would like to thank Costis Papageogakis, Mukund Rangamani and Maximilian Schmidt-Somerfeld for helpful discussions. PR is supported by the STFC studentship grant ST/F007698/1.

\end{document}